\documentclass[fleqn,10pt]{article}
\usepackage[utf8]{inputenc}
\usepackage[T1]{fontenc}

\usepackage{amsfonts}
\usepackage{cite}
\usepackage{graphicx}
\usepackage{dcolumn}
\usepackage{bm}
\usepackage{amsmath}
\usepackage{amssymb}
\usepackage{hhline}
\usepackage[normalem]{ulem}
\usepackage{authblk}

\usepackage{color}

\usepackage{hyperref}
\hypersetup{colorlinks=true,citecolor={blue},linkcolor={blue},urlcolor={blue}}

\newcommand{\clr}[1]{{\color{black} #1}}

\begin{document}

\title{Stochastic circular persistent currents of exciton polaritons}

\author{J.~Borat$^{1,2}$, Roman~Cherbunin$^{3}$, Evgeny Sedov$^{1,2,3,4,*}$, 
Ekaterina~Aladinskaia$^{3}$, Alexey~Liubomirov$^{3}$, Valentina~Litvyak$^{3}$, Mikhail~Petrov$^{3}$, Xiaoqing Zhou$^{1,2}$, Z.~Hatzopoulos$^5$,
Alexey Kavokin$^{1,2,4,6}$, P.~G.~Savvidis$^{1,2,5,7}$ \\
\small
$^1$Key Laboratory for Quantum Materials of Zhejiang Province, School of Science, Westlake University, 18 Shilongshan Rd, Hangzhou 310024, Zhejiang, China\\
$^2$Institute of Natural Sciences, Westlake Institute for Advanced Study, 18 Shilongshan Road, Hangzhou, Zhejiang Province 310024, China\\
$^3$Spin Optics Laboratory, St. Petersburg State University, Ulyanovskaya 1, St. Petersburg 198504, Russia\\
$^4$Vladimir State University named after A. G. and N. G. Stoletovs, Gorky str. 87, Vladimir 600000, Russia\\
$^5$FORTH-IESL, P.O. Box 1527, 71110 Heraklion, Crete, Greece\\
$^6$Moscow Institute of Physics and Technology, Institutskiy per., 9, Dolgoprudnyi, Moscow Region, 141701, Russia\\
$^7$Department of Materials Science and Technology, University of Crete, P.O. Box 2208, 71003 Heraklion, Crete, Greece\\
\small{*evgeny\_sedov@mail.ru}
}

\maketitle

\begin{abstract}
We keep track of the orbital degree of freedom of an exciton polariton condensate, confined in an optical trap, and reveal the stochastic switching of persistent annular polariton currents in the pulse-periodic excitation regime.
In an elliptic trap, the low-lying in energy polariton current states are inherent in a two-petalled density distribution and swirling phase.
In the stochastic regime, the averaged over multiple excitation pulses density distribution gets homogenised in the azimuthal direction, while the weighted  phase extracted from interference experiments experiences two compensating each other jumps, when varying around the center of the trap.
Breaking the reciprocity of the system with a supplemental control optical pulse makes it possible to switch the system from the stochastic regime to the deterministic regime of an arbitrary polariton circulation.
\end{abstract}

\thispagestyle{empty}

\section*{Introduction}

Recently, much attention in polaritonics has been given to the orbital degree of freedom.
Exciton polaritons, eigenmodes of optical microcavities strongly coupled to excitons in embedded semiconductor quantum wells (QWs)~\cite{kavokinBook2017}, form macroscopic states of exciton-polariton condensates that behave like a superfluid liquid~\cite{NatPhys5805,NatPhys13837}.
Flows of polaritons within the condensate state endow the latter with the nonzero orbital angular momentum (OAM).
Polariton condensates in annular~\cite{PhysRevB97195149,ACSPhoton71163,PhysRevResearch3013072,PNAS1122676} and pot-shaped~\cite{Optica8301,PhysRevA99033830} traps, traps of complicated shape~\cite{PhysRevLett113200404,PhysRevX5011034}, trap chains and clusters~\cite{PhysRevB95235301,NatureMaterials161120,NatCommun31243} have been considered for the study of and manipulation by their OAM.
Such attention to this problem is justified by the broad prospects for the use of orbital degree of freedom for quantum and classical information storage and processing~\cite{Nature412313,PhysRevLett89240401, PhysLettA3811858,PhysRevResearch3013099} as well as for optical communications~\cite{AdvOptPhoton766,PhotonRes4B14}.

Polariton vortices are the most prominent representatives of polariton condensate states with nonzero OAM. 
The spontaneous formation of polariton vortices (in the form of vortex-antivortex pairs and clusters) has been extensively studied~\cite{NatPhys4706,PhysRevB88201303,PhysRevLett128237402}.
A separate area of research is the excitation of vortices with predefined OAM (or, equivalently, the direction and distribution of the polariton flow density).
Among the used approaches are the resonant excitation scheme and resonant imprinting of OAM~\cite{NatPhys6527,PhysRevLett104126402,PhysRevB90014504}, 
engineering of the effective complex trapping potential~\cite{PhysRevLett113200404,PhysRevB97195149,ACSPhoton71163,PhysRevResearch3013072} under the incoherent excitation,
and the ill-understood direct transfer of OAM from the non-resonant optical pump beam~\cite{PhysRevLett122045302}.
Incoherent control of polariton vortices was reported in~\cite{PhysRevB93035315} for short-lived polaritons characterized by a lifetime of units of picoseconds. 
Such polaritons are unable to undergo long-range ballistic behavior.
They form macroscopic coherent states predominantly under the pump spot, so the gain-induced trapping is realized for such polaritons~\cite{PhysRevLett104126403,PhysRevB94134310,PhysRevResearch3013099}.

In our paper, we study both experimentally and numerically formation of persistent polariton currents in an optically induced elliptical pot-shaped trap created in a planar microcavity.
In the experiment, we demonstrate controllable non-resonant excitation of polariton condensates supporting internal persistent polariton flows. 
Their two-petal shape, contrary to expectation, does not indicate the formation of a phase-locked standing wave (see, e.g.,~\cite{PNAS1118770,LSA1045}), but coexists with its phase, swirling around the center of the trap.
We demonstrate the stochastic switching between two orthogonal polariton current states under the pulse-periodic nonresonant optical excitation.

Polariton currents resulting in nonzero OAM of the polariton condensate are reflected in a twisted wavefront of its photoluminescence (PL), which makes it possible to judge the polariton flows from the phase distribution of PL.
To reveal circulation of the polariton condensate phase, we use the interferometry measurements~\cite{PhysRevB97195149,ACSPhoton71163}.
We use the Mach-Zehnder interferometer with the spherical reference wave.

\section*{Results}

\subsection*{Experiment} 
\begin{figure}[tb!]
\begin{center}
\includegraphics[width=0.6\linewidth]{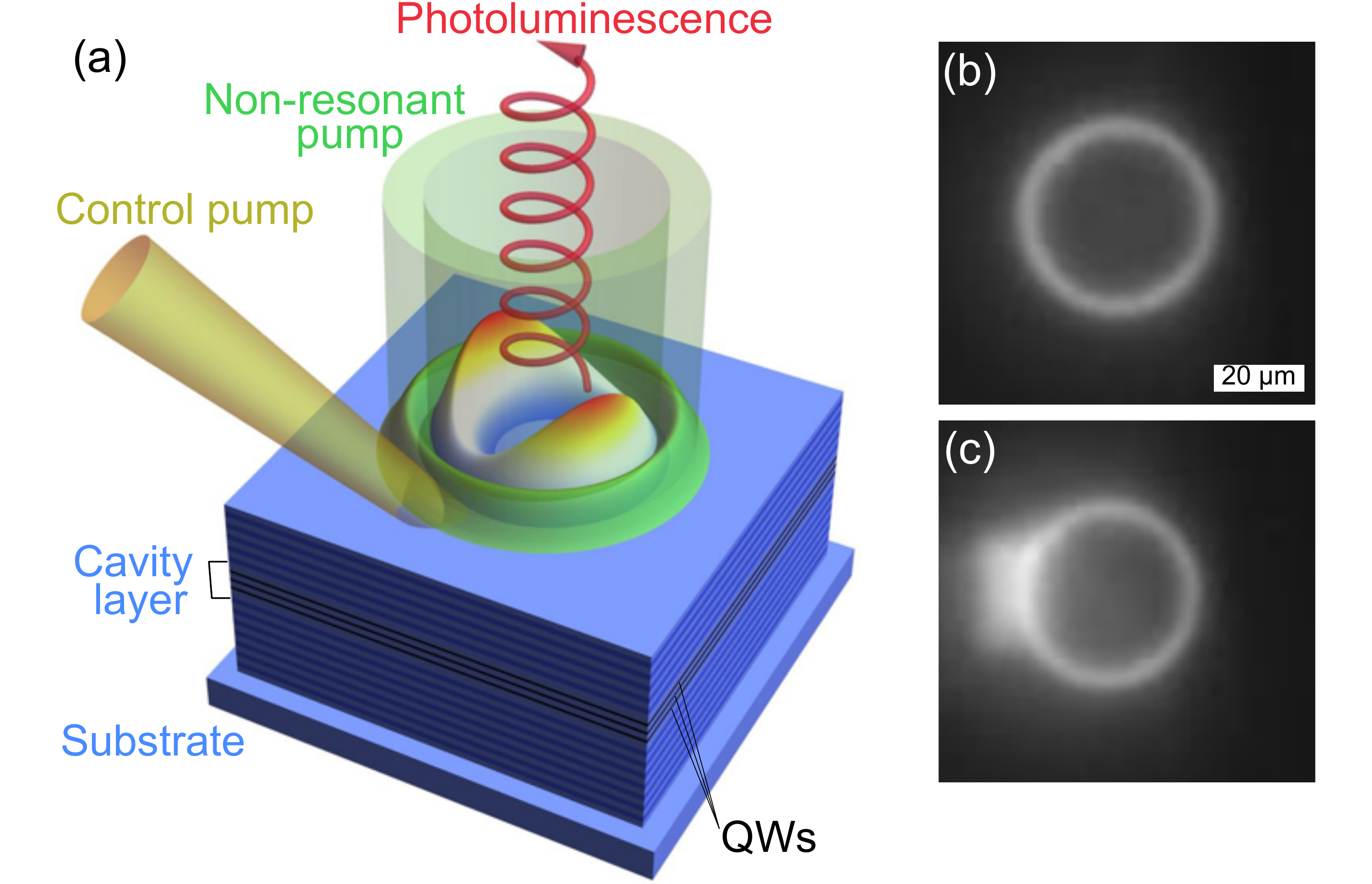}
\end{center}
\caption{ \label{FIG_SCHEME}
(Color online)
(a) Schematic of excitation of the polariton condensate with a nonresonant annular optical pump in a planar microcavity with embedded quantum wells.
Luminescence of the sample under the non-resonant optical pump below the polariton condensation threshold 
in the absence (b) and in the presence (c) of the supplemental control pulse.
(Brightness on the panels increased by 50\% for clarity.)}
\end{figure}

Schematically the creation of a polariton condensate is illustrated in Fig.~\ref{FIG_SCHEME}(a).
The condensate is created in planar optical microcavity with embedded quantum wells by the nonresonant (at the upper Bragg mode of the microcavity) optical pump of annular shape (with radius of about $19 \, \mu \text{m}$) with a weak ellipticity in the pulse-periodic regime with pulse duration of 1~ps and interpulse interval of 13~ns.
See details of the sample under the study in Methods.
Figure~\ref{FIG_SCHEME}(b) shows luminescence from the sample at the pump power considerably below the polariton condensation threshold, which gives an idea of the shape of the pump.
Pumping creates a reservoir of incoherent excitons, which spatial distribution replicates the shape of the pump.
The reservoir acts as a source of polaritons for the condensate feeding it via stimulated relaxation processes.
At the same time, due to the repulsive polariton-exciton interaction, the reservoir plays a role of the trapping potential for polaritons. 
The emerging polaritons in the structure live long enough (with lifetime estimated as tens of picoseconds) to go down the potential hill to the bottom of the trap and occupy eigenstates of the trap.
Due to the dissipative nature of polaritons and to the presence of the pumping, the occupied eigenstate does not have to be the ground state of the trap, but the state with the most advantageous balance of distributed in the microcavity plane losses and gain~\cite{PhysRevB92035305,PhysRevB97235303}.
The latter is determined by the overlapping of the polariton condensate wave function and the cloud of the reservoir excitons.

\begin{figure}[t!]
\begin{center}
\includegraphics[width=\linewidth]{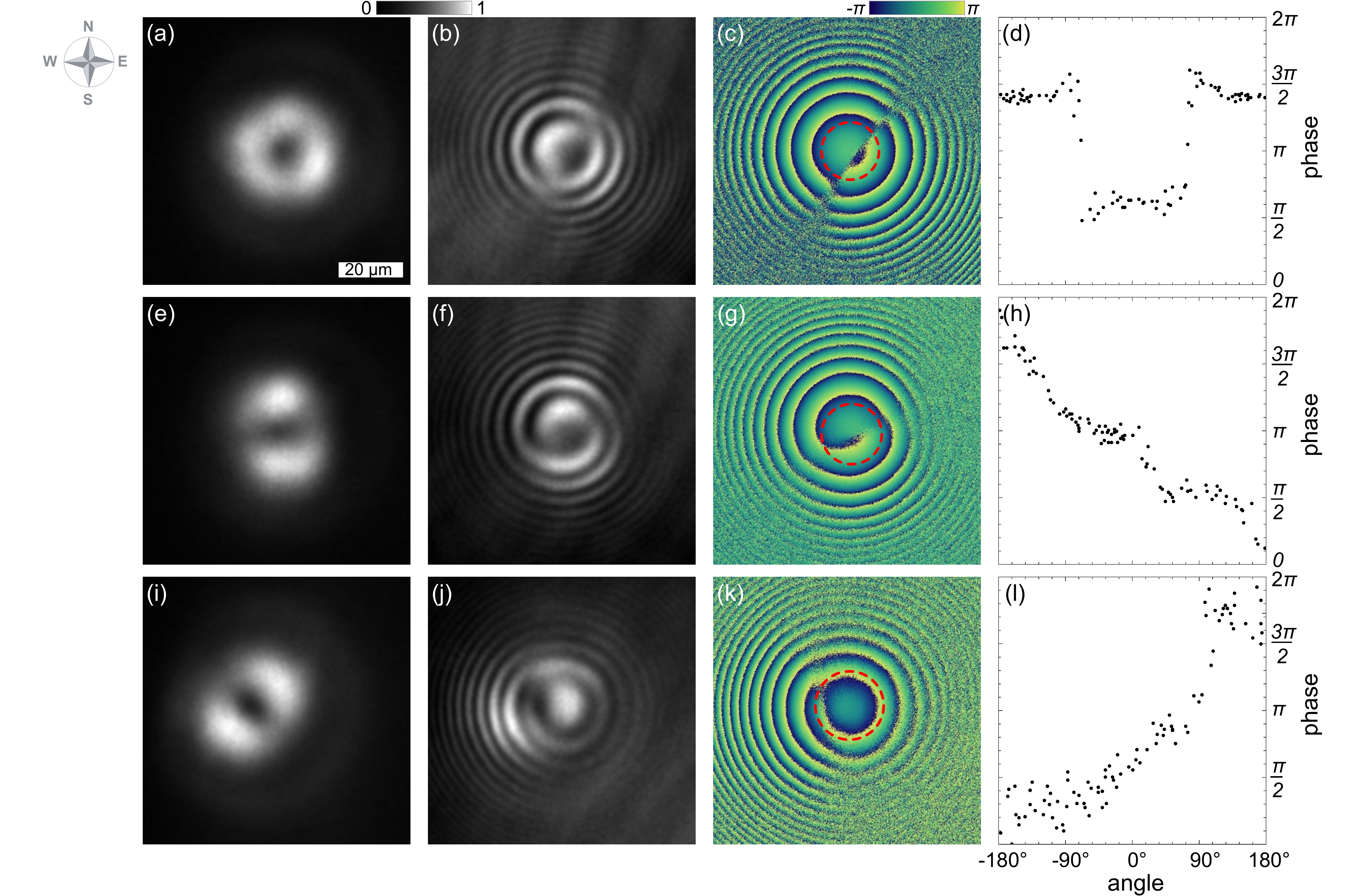}
\end{center}
\caption{ \label{FIG_EXP}
(Color online)
Observation of the polariton condensates in the annular trap.
From left to right: the density distribution of the polariton condensate, the interferometry image, the relative phase map of the condensate, and phase variation along the red dashed circumference in the panel to the left.
Panels (a)--(d) illustrate the condensate in the stochastic switching regime.
Panels (e)--(h) and (i)--(l) illustrate the condensates with cw and ccw polariton currents, respectively. 
Angle $0 ^{\circ}$ in right panels was chosen arbitrary to fit without gaps the phase in the range from 0 to $2\pi$.
}
\end{figure}

\begin{figure}[t!]
\begin{center}
\includegraphics[width=\linewidth]{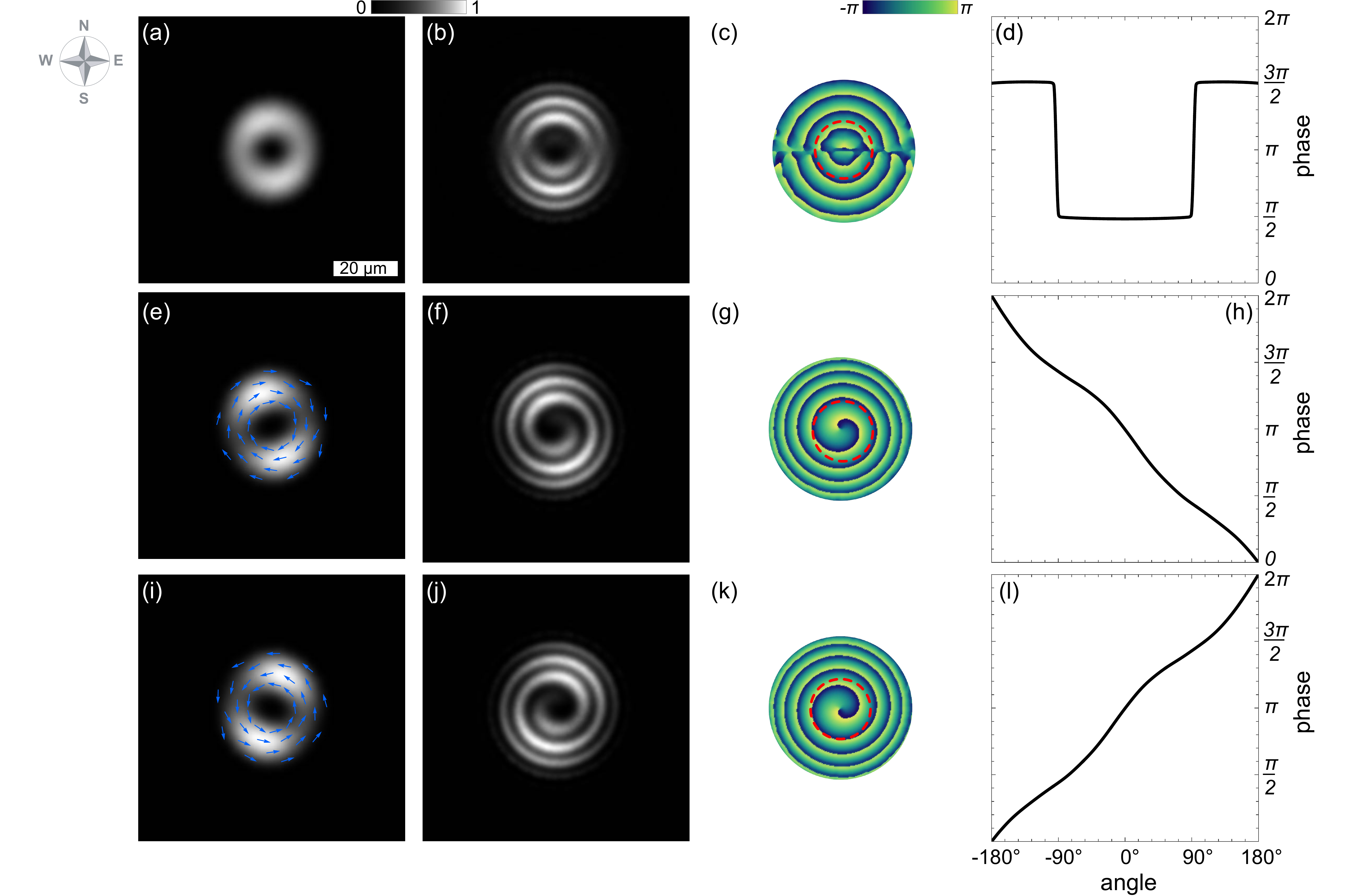}
\end{center}
\caption{ \label{FIG_SIMUL}
(Color online)
Simulation of the polariton condensates in the annular trap.
Meaning of the panels in this figure is the same as that in Fig.~\ref{FIG_EXP}.
Blue arrows in panels~(e) and (i) show the vector field of the current density $\mathbf{J}$.
Values of the parameters used for simulations are given in~Methods.
}
\end{figure}

In the first experiment, we optically excite the polariton condensate and detect its PL spatial distribution, averaged in time, see Fig.~\ref{FIG_EXP}(a).
The polariton condensate has a closed, close to annular shape weakly modulated in the azimuthal direction.
We observe interference of the condensate PL with a spherical reference wave, see the interferometry image in Fig.~\ref{FIG_EXP}(b).
The coherent spherical wave was obtained by magnifying the part of the condensate PL and guiding it through a converging lens distant from the image plane by more than its focal length.
The interference fringes represent concentric rings twice split in two nearly diametrically opposed points, so that the half-rings with maximum intensity (white) change over the half-rings with minimum intensity (black).
To extract the phase of the condensate, we use four interferometry images, $I(\mathbf{r}, \Delta \phi)$, measured at different phase delays in the interferometer, $\Delta \phi = 0, \pi/2, \pi$ and $3 \pi /2$.
The condensate phase relative to the phase of the reference beam then can be calculated as~\cite{ApplOpt223421}
\begin{equation}
\label{EqPhaseExtr}
\Phi (\mathbf{r}) \clr{+} \Phi _{\text{ref}} ({r}) = \clr{\tan^{-1}} \left[ \frac{ I (\mathbf{r}, 3 \pi/2) - I (\mathbf{r}, \pi/2) }{ I (\mathbf{r}, 0) - I (\mathbf{r},\pi)} \right].
\end{equation}
The extracted phase distribution in the microcavity plane of the observed PL is shown in Fig.~\ref{FIG_EXP}(c) supplemented with the phase variation around the center of the trap along the top of the ridge of the condensate in~Fig.~\ref{FIG_EXP}(d).
As one can see, the extracted phase twice jumps (by $-\pi$ and $+\pi$) when going along a closed circular loop around the trap, herewith it remains constant beyond the jumps.
Such peculiarity would allow us to expect that the condensate possesses dips in the density distribution around the phase jumps.  
However, this expectation does not match the observations in Fig.~\ref{FIG_EXP}(a) where the density is nearly homogeneous in azimuthal direction.
This discrepancy allows us to conclude that the observed PL characterizes averaging over multiple realizations of condensates with arbitrary phases, rather than a single condensate with a given phase distribution.

The elliptic trap possesses the axial symmetry, so the counterclockwise (ccw) and clockwise (cw) directions keep to be equal.
This implies the equiprobable occurrence of ccw and cw currents of polaritons in the condensate under each pulse of the pump.
Averaging over multiple pulses gives the interferometry image that accumulates phase peculiarities inherent to each condensate realization during the observation.

To distinguish the current polariton states contributing to PL in Fig.~\ref{FIG_EXP}(a), we make the system chiral breaking equivalence of the ccw and cw directions.
To this end, we supplement each annular pump pulse with a weak control pulse in the form of a stripe that overlaps with the annulus, and the position of the stripe can be arbitrarily changed to modify the shape of the effective optically induced potential.
Figure~\ref{FIG_SCHEME}(c) shows the sub-threshold PL from the sample in the presence of the supplemental control pulse.

The middle (panels(e)--(h)) and lower (panels(i)--(l)) rows in Fig.~\ref{FIG_EXP} illustrates two polariton condensates obtained at the control pulses shifted to the southwest and northwest of the center of the annular trap, respectively.
Both condensates possess a dumbbell shape, and their axes of symmetry are inclined with respect to each other.
The key peculiarity of such states, that differs them from the conventional first excited states of a 2D pot-shaped (in particular, harmonic) trap, is that they are vortical states with internal circular polariton currents.
We make such a conclusion based on the observation of interference of PL with the spherical wave, see Figs.~\ref{FIG_EXP}(f) and~\ref{FIG_EXP}(j). 
Clear single-armed spirals indicate the presence of vortices characterized by topological charges $m=-1$ and $+1$, respectively.
The extracted phase distributions in Figs.~\ref{FIG_EXP}(g)--\ref{FIG_EXP}(h) and~\ref{FIG_EXP}(k)--\ref{FIG_EXP}(l) showing the phase change in one turn around the center of the trap by $-2\pi$ and $+2\pi$, respectively, confirm our statement. 
We treat the observed vortex polariton condensate states as the contributors to PL and interferometry image illustrated in panels (a) and (b) of Fig.~\ref{FIG_EXP}.

\subsection*{Simulations}
To support our claims about the observations, we accompany our experiment with numerical simulations of behavior of the polariton condensate in the optically induced annular pump.
See details of the numerical model in Methods.
Results of our simulations illustrated in Fig.~\ref{FIG_SIMUL} successfully qualitatively reproduce the experimental observations in Fig.~\ref{FIG_EXP}.
In Figs.~\ref{FIG_SIMUL}(a) and \ref{FIG_SIMUL}(b), we show the density distribution and interference image averaged over two different realizations of the polariton condensate with  vortices of opposite circulation directions.
The phase distribution and the azimuthal phase variation in Figs. ~\ref{FIG_SIMUL}(c) and \ref{FIG_SIMUL}(d) were extracted from Fig.~\ref{FIG_SIMUL}(b) using the extended Fourier-transform method for closed-fringe patterns, described in~\cite{PhysRevB97195149}.
Both the shift of the interference fringes and the mutually compensating jumps with a close to homogeneous azimuthal density distribution take place in the images.
We should mention that the increase of number of realizations used for averaging from two to any arbitrary large number does not change the presented picture.

The middle, (e)--(h), and lower, (i)--(l), panels in Fig.~\ref{FIG_SIMUL} illustrate polariton condensate vortex states contributing to the stochastic switching regime in panels (a)--(d).
The simulations reproduce the formation of the dumbbell-shaped condensates with the swirling phases, nearly linearly varying around the vortex core.
Polariton currents characterized by the vector field of the current density $\mathbf{J} = \text{Im} \left( \Psi ^* \nabla \Psi \right)$ are shown in panels~\ref{FIG_SIMUL}(e) and~\ref{FIG_SIMUL}(i) by blue arrows.

\section*{Discussion}

In this work, we have demonstrated the regime of stochastic switching of polariton condensate vortices (circular currents) in the complex annular trapping potential.
In the PL measurements, the averaged in time intensity distribution takes a ring shape weakly modulated in the azimuthal direction.
In the interferometry measurements with the spherical reference wave, the averaged interference fringes represent concentric rings split in two nearly diametrically opposed points.
Making the trapping potential chiral by applying the supplemental weak control pump pulse, we can convert the system to the regime of the deterministic circulation, in which polariton vortices emerge with the circulation direction depending on the position of the control pulse. 

We should  specify, that since in the stochastic switching regime, in Figs.~\ref{FIG_EXP}(b) and~\ref{FIG_SIMUL}(b) averaging over intensity takes place, the extracted phase distributions in Figs.~\ref{FIG_EXP}(c) and~\ref{FIG_SIMUL}(c) do not characterize the phase properties of either of the contributing condensates.
Herewith, the interferometry measurements make it possible to clearly distinguish between the stochastic and deterministic regimes of the behavior of the polariton condensate without deep analysis.
This favorably distinguishes the interference method from other approaches to the revealing phase properties of polariton condensates, such as the OAM sorting method~\cite{PhysRevLett105153601,OptLett2632248} 
and the machine-learning-based approach combining  dimensionality reduction and linear regression techniques~\cite{MachLearn2022}. 

In the experiments on the deterministic circulation of polaritons, the annular pump pulse and the weak control pulse arrive at the sample almost simultaneously.
However, there is often a delay of one pulse relative to another.
We point out that the formation of circular polariton currents of a given direction is insensitive to which of the two pulses comes first, at least, in the interval of delays of $\pm500$~ps.
This observation allows us to conclude that lifetime of the optically induced reservoir excitons contribution to the trapping potential is of the order of hundreds of picoseconds.

In all our experiments, PL from the condensate region was observed as long as the optical pump was switched on.
Once established, both PL and interferometry images remained unchanged until the pump is turned off or the pump regime is changed.
Remarkably, the positions of the breaks of the interference fringes in the stochastic switching regime remained unchanged in different experiments, after tuning the pump off and on again.
This may be understood having in mind that the fringes are governed by the phase difference between the condensate PL and the reference wave.
This phase depends on the position of the spot of the condensate used for obtaining the reference wave and on the phase delay between the arms of the interferometer, which do not change during the experiment.
Retaining the reference wave phase and the interferometer delay unchanged is also responsible for the fact that in the averaged interferometry images we observe clearly distinguished interference fringes rather than a uniformly illuminated spot. This confirms that the phase difference with the reference wave for the polariton condensate vortices of either direction is kept the same.
Finally, the observed stochasticity may be of either classical or quantum nature. The present experiments do not allow us to rule out any of these two options.

\section*{Methods}

\subsection*{Sample details}
The sample under study is a planar $5 \lambda/2$ AlGaAs microcavity with an embedded set of GaAs QWs, sandwiched between two distributed Bragg reflectors, formed of 31 (top) and 35 (bottom) AlGaAs/AlAs pairs of layers.
The bottom mirror is grown on the GaAs substrate.
The  cavity quality factor is estimated as $Q > 10^{4}$.
The detuning of the QW exciton resonance from the cavity mode varies over the sample from $-20$~meV to $+3$~meV.
The strong exciton-photon coupling regime is supported by the sample with the Rabi splitting of 9.2~meV.

\subsection*{Experimental setup}

The experimental setup has a fairly standard schematic. The sample is placed at the cold finger of a low-vibration closed-cycle exchange-gas cryostat and cooled down to $T \approx 6$~K. As the pumping source, we use a ps-pulsed Ti:Sapphire laser tuned to a second dip in the reflectance spectrum of the high-energy DBR side at $\lambda_\mathrm{pump} = 751$~nm. The laser beam is split onto two separate beams with the intensity ratio of 8:92. The stronger-power beam follows through the acousto-optic modulator, used for a controllable reduction of the beam intensity, and the MEMS-based spatial light modulator to form a ring-shape trapping-pulse beam. The lower-power beam is guided to a $300$~mm mechanical delay stage and further spatially modulated by a slit-based spatial filter to form a small-intensity control-pulse beam that can be switched on/off by a mechanical shutter. The trapping and control beams are then focused on the sample into a spot of diameter of about $20$~$\mu$m in diameter by using a $F=4$~mm microscope objective. A relatively high numerical aperture of the imaging lens, $\text{\textit{NA}}=0.42$, allows us to perform both the real-space and the reciprocal-space analysis of the trapped polariton condensates. The PL is collected in the reflection geometry by the same objective, separated from a back-reflected laser light by a long-pass interference filter, and further split into two beams. To resolve a condensate emission in the reciprocal space, the first PL beam is used to refocus the back focal plane of the microscope objective to a slit of the $0.5$~m imaging spectrometer. The second PL beam is analyzed in a real space by using the Mach-Zehnder interferometer where we place a $F = 60$~mm lens in one arm, i.e., a part of the condensate glow is used as a reference spherical wave. The second arm of the interferometer, the object arm, is tuned by a precision mechanical delay line to adjust the optical paths within the range of the first-order correlation function decay. The interference pattern is analyzed with the magnification factor of $\times62.5$ by a thermoelectrically cooled CMOS camera exposed during times from $0.5$ to $2.5$~s. The interferograms are captured multiple times with the small shift of the interferometer path difference to collect four interferograms corresponding to $\Delta\phi = 0$, $\pi/2$, $\pi$, and $3\pi/2$ phase delays. The measured interferogram gives
\begin{equation}
I(\mathbf{r},\Delta\phi) =  I'(\mathbf{r}) + I''(\mathbf{r})\cdot\cos[\Theta(\mathbf{r})+\Delta\phi].
\end{equation}  
Here, $I'(\mathbf{r}) = I_\mathrm{sig} + I_\mathrm{ref}$, $I''(\mathbf{r}) = 2\sqrt{ I_\mathrm{sig} I_\mathrm{ref}}$ where $I_\mathrm{sig}$ and $I_\mathrm{ref}$ are the intensities of the object and reference waves. The relative phase difference, $\Theta(\mathbf{r})$, can be recovered using a four-step phase-reconstruction algorithm~\cite{OptSpec}
\begin{equation}
\Theta(\mathbf{r}) = \tan^{-1} \left[\frac{I(\mathbf{r}, 3\pi/2) - I(\mathbf{r},\pi/2)}{I(\mathbf{r}, 0) - I(\mathbf{r},\pi)}\right].
\end{equation}
The rest parts of the interferogram can be obtained accordingly
\begin{subequations}
\begin{align}
&I'(\mathbf{r}) = [I(\mathbf{r},0) + I(\mathbf{r},\pi)]/2,\\
&I''(\mathbf{r})\cdot\sin\Theta(\mathbf{r}) = [I(\mathbf{r},3\pi/2) - I(\mathbf{r},\pi/2)]/2.
\end{align}
\end{subequations}

To further evaluate the data, in the separate measurement, an object arm of the interferometer was cut by a mechanical shutter to collect only the intensity of the reference wave, $I_\mathrm{ref}$. Suppose the reference beam has a flat intensity over the whole sensor area and the phase exhibits only a slowly-varied parabolic shape, $\Phi_\mathrm{ref}(\mathbf{r})$, centered at the origin of the object wave. In this case, the complex field of the condensate emission, $\Psi(\mathbf{r}) = \|\Psi(\mathbf{r})\|\cdot\exp[i\,\Phi(\mathbf{r})]$, can be evaluated as follows~\cite{ACSPhoton71163}
\begin{subequations}
\begin{align}
	&\|\Psi(\mathbf{r})\| = I'(\mathbf{r}) - I_\mathrm{ref},\\
	&\Phi(\mathbf{r}) = \Theta(\mathbf{r}) - \Phi_\mathrm{ref}(\mathbf{r}),
\end{align}
\end{subequations}
without a need to perform the complex 2D Fourier analysis usually implemented in off-axis digital holography experiments.

\subsection*{The model used for numerical simulations}
For describing evlution of the polariton condensate, we use the generalized Gross-Pitaevskii equation for the polariton wave function $\Psi (t,\mathbf{r})$~\cite{NJPhys14075020,PhysRevLett109216404}:
\begin{equation}
\label{EqGPE}
i \hbar \partial _t \Psi (t,\mathbf{r}) = \left\{ [i \Lambda _0 n_{\text{R}} (t,\mathbf{r}) - 1] \frac{\hbar ^2}{2 M} \nabla ^2 
+ \alpha |\Psi (t,\mathbf{r})|^2 
 + \alpha _{\text{R}} n_{\text{R}} (t,\mathbf{r})
+ \frac{i \hbar}{2} \left[ R n_{\text{R}} (t,\mathbf{r}) - \gamma  \right\} \vphantom{\frac{\hbar ^2}{2 M}}\right]
\Psi (t,\mathbf{r}),
\end{equation}
coupled to the rate equation for the density of the exciton reservoir $n_{\text{R}} (t,\mathbf{r})$:
\begin{equation}
\label{EqResEq}
\partial _t n_{\text{R}} (t,\mathbf{r}) = P (\mathbf{r}) - \left[ \gamma _{\text{R}} + R |\Psi (t,\mathbf{r})|^2 \right] n_{\text{R}} (t,\mathbf{r}).
\end{equation}
In~Eq.~\eqref{EqGPE}, $M$ is the effective mass of polaritons in the microcavity plane, $\alpha$ and $\alpha _{\text{R}}$ are interaction constants of polaritons with each other within the condensate and polaritons with the reservoir excitons, respectively. 
The rightmost square brackets in Eq.~\eqref{EqGPE} are responsible for the balance of gain and losses in the polariton condensate.
$R$ is the stimulated scattering rate from the reservoir to the condensate,
$\gamma$ and $\gamma _{\text{R}}$ are the decay rates of polaritons and reservoir excitons.
The reservoir is excited by the optical pump of the ring shape $P(\mathbf{r}) \propto \exp \left[ -( \sqrt{x^2 + (y/s)^2} - R )^2 /2 w^2 \right]$, where $R$, $w$ and $s$ are the radius, width and ellipticity of the annulus.

In Eq.~\eqref{EqGPE}, in the imaginary part of the kinetic energy term, we take into account the energy relaxation of propagating polaritons due to their interaction with the reservoir excitons~\cite{NJPhys14075020,PhysRevLett109216404}.
The relaxation is the stronger, the more the condensate wave function overlaps with the reservoir.
$\Lambda _0$ is the fitting parameter describing the contribution of the energy relaxation to the polariton behavior.

\subsection*{Values of the parameters}
We take the following values of the parameters for numerical simulations.
The effective mass of polaritons is $M = 4 \cdot 10^{-5} m_{\text{e}}$, where $m_{\text{e}}$ is the free electron mass.
The polariton and exciton decay rates are taken as $\gamma = 0.025$~ps$^{-1}$ and $\gamma _{\text{R}} = 0.01$~ps$^{-1}$, respectively.
The stimulated scattering rate is taken as $\hbar R=0.1\, \text{meV} \, \mu \text{m}^2$.
The nonlinearity coefficients are taken as $\alpha = \alpha_{\text{R}}/2 = 3 \, \mu \text{eV} \, \mu \text{m}^2$.
The fitting parameter of energy relaxation is $\Lambda _0 = 0.015\,\mu \text{m}^2$.
The radius, width and ellipticity of the pump annulus are $R = 19 \, \mu\text{m}$, $w = 4 \, \mu\text{m}$ and $s=1.08$, respectively.

\bibliographystyle{unsrt}
\bibliography{stochcurrsBibl}

\section*{Acknowledgements}

The experiment was carried out in the Spin Optics Laboratory at the Saint-Petersburg State University (Grant No. 91182694).
A.K. and P.S. acknowledge the support of Westlake University, Project 041020100118 and Program 2018R01002 funded by Leading Innovative and Entrepreneur Team Introduction Program of Zhejiang Province of China.
A.K.~acknowledges support from the Moscow Institute of Physics and Technology under the Priority 2030 Strategic Academic Leadership Program.
Numerical simulations were supported by the RF Ministry of Science and Higher Education under Agreement No. 0635-2020-0013 and by the RF Presidential Grant for state support of young scientists (No. MK-4729.2021.1.2).

\end{document}